\documentclass[a4paper,11pt]{article}
\usepackage{pos}
\usepackage{ulem}
\usepackage{color}
\usepackage{amsfonts}
\usepackage{mathrsfs}
\usepackage{bm}
\usepackage[dvipdfmx]{graphicx}
\usepackage{longtable}

\title{Towards the continuum limit of nucleon form factors at the physical point using lattice QCD}
\ShortTitle{Towards the continuum limit of nucleon form factors....}

\author*[\dag, a.b]{Ryutaro Tsuji}
\author[b]{Yasumichi Aoki}
\author[c]{Ken-Ichi Ishikawa}
\author[d]{Yoshinobu Kuramashi}
\author[a]{Shoichi Sasaki}
\author[d]{Eigo Shintani}
\author[e.d]{Takeshi Yamazaki}
\note[\dag]{Present adress is \textit{RIKEN Center for Computational Science, 650-0047, Kobe, Japan}.}
\affiliation[]{\normalsize{\bf \sffamily \hspace{50mm}(PACS Collaboration)}}

\affiliation[a]{Department of Physics, Tohoku University,\\
  980-8578,Sendai, Japan}

\affiliation[b]{RIKEN Center for Computational Science,\\
  650-0047, Kobe, Japan}

\affiliation[c]{Core of Research for the Energetic Universe,\\
    Graduate School of Advanced Science and Engineering, Hiroshima University\\
  739-8526, Higashi-Hiroshima, Japan}

\affiliation[d]{Center for Computational Sciences, University of Tsukuba,\\
  305-8577, Tsukuba, Japan}

\affiliation[e]{Faculty of pure and Applied Sciences, University of Tsukuba,\\
  305-8571, Tsukuba, Japan}

\emailAdd{tsuji@nucl.phys.tohoku.ac.jp}

\abstract{
    We present results for the axial charge and root-mean-square (RMS) radii of the nucleon obtained from 2+1 flavor lattice QCD at the physical point with a large spatial extent of about 10 fm. Our calculations are performed with the PACS10 gauge configurations generated by the PACS Collaboration with the six stout-smeared O(a) improved Wilson-clover quark action and Iwasaki gauge action at $\beta$ = 1.82 and 2.00 corresponding to lattice spacings of 0.085 fm and 0.063 fm respectively. We first evaluate the value of $g_{A}/g_{V}$, which is not renormalized in the continuum limit and thus ends up with the renormalized axial charge. Moreover, we also calculate the nucleon elastic form factors and determine three kinds of isovector RMS radii such as electric, magnetic and axial ones at the two lattice spacings. We finally discuss the discretization uncertainties on renormalized axial charge and isovector RMS radii towards the continuum limit.
}

\FullConference{
 The 39th International Symposium on Lattice Field Theory, LATTICE2022
  8th-13th August, 2022
  Bonn University, Bonn, Germany
}

\begin{document}
\maketitle

\section{Introduction}
\label{Sec_intro}
In the standard model of modern particle physics, protons and neutrons, known as nucleons, are composite particles of quarks and gluons, and the interaction among them is formulated as Quantum Chromodynamics (QCD).
This indicates that the nucleon has a non-trivial structure due to the complex dynamics of QCD.
One of the topics that have recently come under the spotlight is the ``size'' of the nucleon such as electric $(\langle r_E^2 \rangle)$, magnetic $(\langle r_M^2 \rangle)$ and axial $(\langle r_A^2 \rangle)$ radii, which can be extracted from the corresponding form factors~\cite{Meyer:2022mix}.

The experimental measurement of the electric radius has a significant puzzle, which is known as the proton radius puzzle~\cite{Bernauer:2020ont}. This puzzle arises from the experiments and is not solved yet. As for the magnetic radius, there is still large uncertainty. It has been reported that the experiments give a different behavior of the magnetic form factor depending on the parametrization~\cite{Borah:2020gte, Bradford:2006yz}. 
This uncertainty leads to the fact that the magnetic radius is not well determined experimentally yet. There is also some tension in the axial radius between the experiment and lattice QCD~\cite{Meyer:2022mix}. This tension causes a large discrepancy in the neutrino-nucleon scattering amplitude, which has an important role in dark matter search experiments~\cite{Monroe:2007xp}.

The lattice QCD community has also computed the size of the nucleon. 
Towards the high-precision determination by lattice QCD, the major sources of uncertainties are identified as follows: statistical noise, excited-state contaminations, model dependences for extracting radii from data, finite-size effects, and chiral-continuum extrapolation~\cite{Djukanovic:2021qxp}. 
Recent lattice QCD calculations have succeeded in reproducing the results being consistent with the experiments~\cite{Jang:2019vkm, Djukanovic:2021cgp, RQCD:2019jai, Park:2021ypf}. However, they are not enough precise to draw a firm conclusion on the above mentioned issues.

This work presents the preliminary result of our study.
In our previous work~\cite{Shintani:2018ozy}, although most of the uncertainties are handled, the discretization uncertainty is not examined yet. 
Therefore we calculate on the second PACS10 ensemble in order to study the discretization uncertainties of the nucleon form factors.

\section{Method}
\label{Sec_method}
We calculate the electric and magnetic form factors, $G_E(q^2)$, $G_M(q^2)$ and the axial form factor $F_A(q^2)$.
The first two are relevant for the electron-nucleon scattering experiment, while the latter provides crucial information for building neutrino-nuclear cross-section from neutrino-nucleon scatterings.

We simply focus on the isovector quantities, where the disconnected contributions are canceled by each other under the exact SU(2) isospin symmetry~\cite{Sasaki:2003jh}. Therefore the isovector electric and magnetic form factors are given by the combination of proton's and neutron's form factors,
\begin{align}
    G^{v}_l(q^2) = G^{p}_{l}(q^2)-G^{n}_{l}(q^2),
    \quad l=\{E,M\},
\end{align}
%
\text{color}{which} are used for comparison with experimental values. On the other hand, the axial form factor can be directory compared with phenomenological values provided by the neutron $\beta$ decay. 
As for the axial form factor, the axial vector coupling, $g_A = F_A(q^2=0)$, is experimentally well determined as $g_A=1.2756(13)$~\cite{ParticleDataGroup:2020ssz}. Therefore, we also calculate this quantity as a good reference for checking calculation accuracy.

The nucleon 2-point function with the nucleon interpolating operator located at either smeared (\textit{S}) or local (\textit{L}) sources $(t_{\rm src})$, and local sink $(t_{\rm sink})$ is constructed as

\begin{align}
    \label{eq:two_pt_func}
    C_{XS}(t_{\rm sink}-t_{\rm src}; \bm{p})
    =
    \frac{1}{4}\mathrm{Tr}
    \left\{
        \mathcal{P_{+}}\langle N_X(t_{\mathrm{sink}}; \bm{p})
        \overline{N}_S(t_{\mathrm{src}}; -\bm{p})  \rangle
    \right\}\ \mathrm{with}\ X=\{S,L\},
\end{align}

\noindent
where $\mathcal{P_{+}}=(1+\gamma_4)/2$, which can eliminate the unwanted contributions from the negative-parity state for $|\bm{p}|=0$~\cite{Sasaki:2001nf}.
In this study, the smeared operators are constructed by
the exponentially smeared quark operators, so as to maximize
overlap with the nucleon ground state as

\begin{align}
    \label{eq:interpolating_op}
    N(t,\vec{p})&
    =
    \sum_{\vec{x}\vec{x_1}\vec{x_2}\vec{x_3}}
    \mathrm{e}^{-i\vec{p}\cdot\vec{x}}\varepsilon_{abc}
    \left[
        u^{T}_{a}(t,\vec{x_1})C\gamma_5d_b(t,\vec{x_2})
    \right]
    u_c(t,\vec{x_3})
    \times 
    \Pi_{i=1}^{3}
    \phi(\vec{x_i}-\vec{x}),
\end{align}
where the smearing function $\phi(\vec{x_i}-\vec{x}) = A\mathrm{e}^{(-B|\vec{x_i}-\vec{x}|)}$ is parameterized with two parameters $(A, B)$. For simplicity, $\vec{x_1} = \vec{x_2} = \vec{x_3}$ is chosen.

The nucleon isovector form factors can be extracted from the nucleon 3-point functions,

\begin{align}
    \label{eq:three_pt_func}
    C^{k}_{\mathcal{O}_\alpha}(t; \bm{p}^{\prime}, \bm{p})
    =
    \frac{1}{4} \mathrm{Tr}
    \left\{
        \mathcal{P}_{k}\langle N(t_{\mathrm{sink}}; \bm{p})
        J^{\mathcal{O}}_{\alpha}(t; \bm{q}=\bm{p}-\bm{p}^{\prime})
        \overline{N}(t_{\mathrm{src}}; -\bm{p})  \rangle
    \right\},
\end{align}

\noindent
where $\mathcal{P}_k$ is a projection operator as $\mathcal{P}_t=\mathcal{P}_{+}$ and $\mathcal{P}_{53}=\mathcal{P}_{+}\gamma_5\gamma_3$, and $J^{\mathcal{O}}_{\alpha}$ represents isovector local current operators as $J^{\mathcal{O}}_\alpha = \bar{u}\mathcal{O}_\alpha u - \bar{d}\mathcal{O}_\alpha d$ with $\mathcal{O}_\alpha = \gamma_\alpha, \gamma_\alpha\gamma_5$ for the vector $(V_\alpha)$ and axial-vector $(A_\alpha)$ currents, respectively. 
In a conventional way to extract the form factors, we take an appropriate combination of 2-point function~(\ref{eq:two_pt_func}) and 3-point function~(\ref{eq:three_pt_func}), 

\begin{align}
\mathcal{R}_{\mathcal{O}_\alpha}^{k}\left(t; \bm{p}^{\prime}, \bm{p}\right) = \frac{C_{\mathcal{O}_\alpha}^{k}\left(t; \bm{p}^{\prime}, \bm{p}\right)}{C_{SS}\left(t_{\mathrm{sink}}-t_{\mathrm{src}}; \bm{p}^{\prime}\right)}
\sqrt{\frac{C_{L S}\left(t_{\mathrm{sink}}-t; \bm{p}\right) C_{SS}\left(t-t_{\mathrm{src}}; \bm{p}^{\prime}\right) C_{L S}\left(t_{\mathrm{sink}}-t_{\mathrm{src}}; \bm{p}^{\prime}\right)}{C_{L S}\left(t_{\mathrm{sink}}-t; \bm{p}^{\prime}\right) C_{S S}\left(t-t_{\mathrm{src}}; \bm{p}\right) C_{LS}\left(t_{\mathrm{sink}}-t_{\mathrm{src}}; \boldsymbol{p}\right)}},
\end{align}

\noindent
which leads to the following asymptotic values,

\begin{align}
    R^t_{V_t}(t;\bm{p})
    &=
    \frac{1}{Z_V}\sqrt{\frac{E_N(p)+M_N}{2E_N(p)}}G_E(q^2),\\
    R^{53}_{Vi}(t;\bm{p})
    &=
    \frac{1}{Z_V}
    \frac{i\varepsilon_{i3k}q_k}{\sqrt{2E_N(p)(E_N(p)+M_N)}}G_M(q^2),\\
    R^{53}_{A_i}(t;\bm{p})
    &=
    \frac{1}{Z_A}
    \sqrt{\frac{E_N(p)+M_N}{2E_N(p)}}
    \left[
        F_A(q^2)\delta_{i3}-\frac{q_iq_3}{E_N(p)+M_N}F_P(q^2)
    \right]
\end{align}

\noindent
under the condition of $t_{\rm sink}\gg t \gg t_{\rm src}$, where the excited-state contaminations are negligible. 
We determine the form factors in the asymptotic region (hereafter denoted as the plateau method).

The RMS radius of a form factor $\mathcal{G}_\mathcal{O}(q^2)$ can be read off from the slope at $q^2=0$,
            
\begin{align}
    \langle r^2_\mathcal{O} \rangle
    =
    -\frac{6}{\mathcal{G_O}(0)}
    \left.
        \frac{d\mathcal{G_O}(q^2)}{dq^2}
    \right|_{q^2=0}
\end{align}

\noindent
with $\mathcal{G_O}=G_E,G_M,F_A$.
In this study, the z-expansion method, which is known as an model independent analysis, is employed~\cite{Boyd:1995cf, Hill:2010yb}. We make a fit to form factors $\mathcal{G_O}(q^2)$ by the following functional form,

\begin{align}
    \label{eq:zexpansion}
    \mathcal{G_O}(q^2)
    =
    \sum_{k=0}^{k_{\mathrm{max}}}c_kz(q^2)^k=
    {c_0+c_1z(q^2)+c_2z(q^2)^2+c_3z(q^2)^3}+\dots,
\end{align}

\noindent
where a new variable $z$ is defined by a conformal mapping from $q^2$ as

\begin{align}
    z(q^2)
    =
    \frac{\sqrt{t_{\rm cut}+q^2}-\sqrt{t_{\rm cut}}}
    {\sqrt{t_{\rm cut}+q^2}+\sqrt{t_{\rm cut}}}
\end{align}

\noindent
with $t_{\rm cut}=4M_\pi^2$ for $G_E$ and $G_M$, or with $t_{\rm cut}=9M_\pi^2$ for $F_A$.
In this study, $k_{\mathrm{max}}=3$ is adopted.

\section{Simulation details}
We mainly use the PACS10 configurations generated by the PACS Collaboration with the six stout-smeared ${\mathscr{O}}(a)$ improved Wilson-clover quark action and Iwasaki gauge action at $\beta=1.82$ and $2.00$ corresponding to the lattice spacings of $0.085$ fm (coarser) and $0.064$ fm (finer), respectively~\cite{Iwasaki:1983iya, Ishikawa:2018jee, Shintani:2019wai, Tsuji:2021bdp}. 
When we compute nucleon 2-point and 3-point functions, the all-mode-averaging (AMA) technique~\cite{Blum:2012uh, Shintani:2014vja} is employed in order to reduce the statistical errors significantly without increasing
computational costs. 
The nucleon interpolating operators defined in Eq.~(\ref{eq:interpolating_op}) are exponentially smeared with $(A, B)=(1.2,0.16)$ for $128^4$ lattice ensemble and $(A,B)=(1.2,0.11)$ for $160^4$ lattice ensemble. As for the 3-point functions, the sequential source method is employed and calculated with $t_{\rm sep}/a=\{10,12,14,16\}$ for $128^4$ lattice ensemble and $t_{\rm sep}/a=\{16,19\}$ for $160^4$ lattice ensemble.

\begin{table*}[ht]
\caption{Summary of simulation parameters 
used in this study.
\label{tab:simulation_details}}
\begin{tabular}{cccccccc}
          \hline        
&  $\beta$ &$L^3\times T$ & $a^{-1}$ [GeV] & $La$ [fm]& $\kappa_{ud}$ & $\kappa_{s}$& $M_\pi$ [GeV] \\
          \hline        
          \hline        
    $128^4$ lattice& 1.82& $128^3\times 128$& 2.3162(44)& 10.9&0.126117 & 0.124902 &  0.135\\
          \hline
    $160^4$ lattice& 2.00& $160^3\times 160$ & 3.1108(70)& 10.1 &0.12584  &0.124925  & 0.138 \\
          \hline
\end{tabular}
\end{table*}

\section{Numerical results}
In this study, we would like to present the preliminary results of $g_{A}/g_{V}$ obtained from the $160^4$ lattice ensemble. Comparing with the results obtained from the $128^4$ lattice ensemble, we will discuss the discretization uncertainty on $g_{A}/g_{V}$ that is not renormalized in the continuum limit, and thus it is associated with the renormalized axial charge. We then next show the preliminary results of nucleon's isovector electric, magnetic and axial RMS radii obtained from the nucleon elastic form factor for each current. We will also discuss the discretization uncertainties on these radii, later.
Before showing these results, we first examine the dispersion relation of the nucleon at both the coarser and finer lattice spacings, and then confirm that the ones observed in our simulations agree well with the relativistic continuum dispersion relation up to our highest momentum transfer. This indicates that the on-shell $O(a)$ improvement works properly at the finite momentum
we used in this study.

\subsection{Ratios of $g_{A}/g_{V}$ from $128^4$ and $160^4$ lattice ensembles}

\begin{figure*}
 \centering
 \includegraphics[width=0.48\textwidth,bb=10 0 725 550, clip]{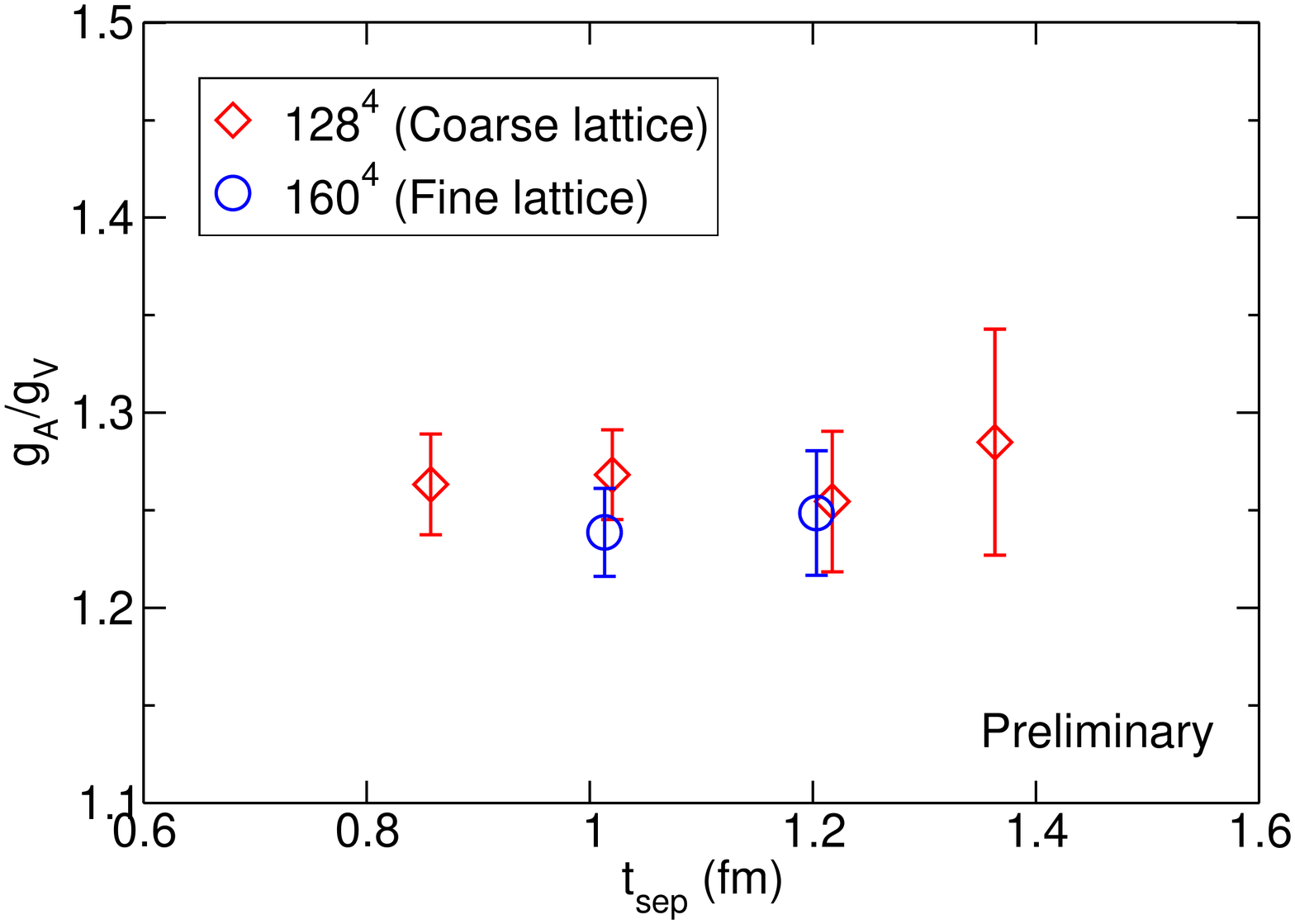}
 \includegraphics[width=0.51\textwidth,bb=10 35 752 725, clip]{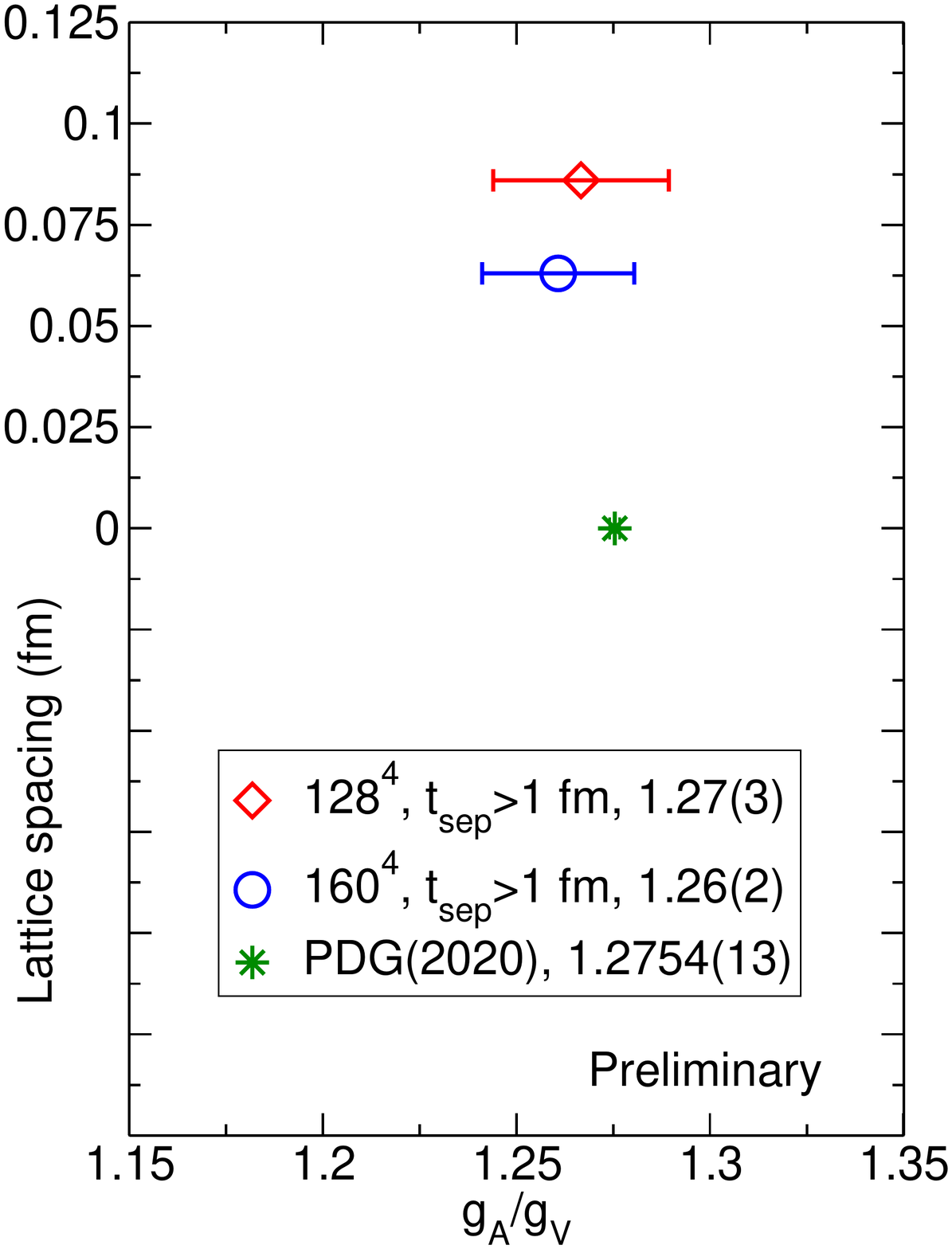}
 \caption{Preliminary results of $g_{A}/g_{V}$ obtained from $128^4$ and $160^4$ lattice ensembles. We show $t_{\mathrm{sep}}$ dependence (left) and lattice discretization dependence (right), respectively.
}
 \label{fig:continuum_esccomp}
\end{figure*}

We first present our preliminary results of $g_{A}/g_{V}$  obtained from both $128^4$ and $160^4$ lattice ensembles, since the renormalization constants are not yet evaluated at the finer lattice spacing.
In the left panel of Fig~\ref{fig:continuum_esccomp}, we show the $t_{\rm sep}$ dependence of the ratios of $g_{A}/g_{V}$,
which are evaluated by the standard plateau method. 
At first glance, the $128^4$ lattice results suggest that the condition of $t_{\rm sep}\ge1$ fm is large enough to eliminate the excited-state contaminations within statistical precision.
The preliminary results of $g_{A}/g_{V}$ from the $160^4$ lattice ensemble, keeping the same condition of $t_{\rm sep}\ge1$ fm, reveal consistent results with the coarser case.

We then examine the discretization uncertainty on $g_{A}/g_{V}$ in the right panel of Fig.~\ref{fig:continuum_esccomp}. For both $128^4$ and $160^4$ lattices, we plot the results obtained from
the combined analysis with the selected data ($t_{\rm sep}/a=\{12, 14, 16\}$ for $128^4$ lattice and $t_{\rm sep}/a=\{16, 19\}$ for $160^4$ lattice) by using the correlated constant fits.
The figure shows that both the coarser and finer results well reproduce the experimental value~\cite{ParticleDataGroup:2020ssz} at the level of the statistical precision of about 2\%.
This indicates that the size of the discretization uncertainties on the renormalized axial charge is less than 2\% at most. 

\subsection{Nucleon elastic form factors and Root-Mean-Square (RMS) radii}

\begin{figure*}
 \centering
 \includegraphics[width=0.32\textwidth,bb=10 20 752 725, clip]{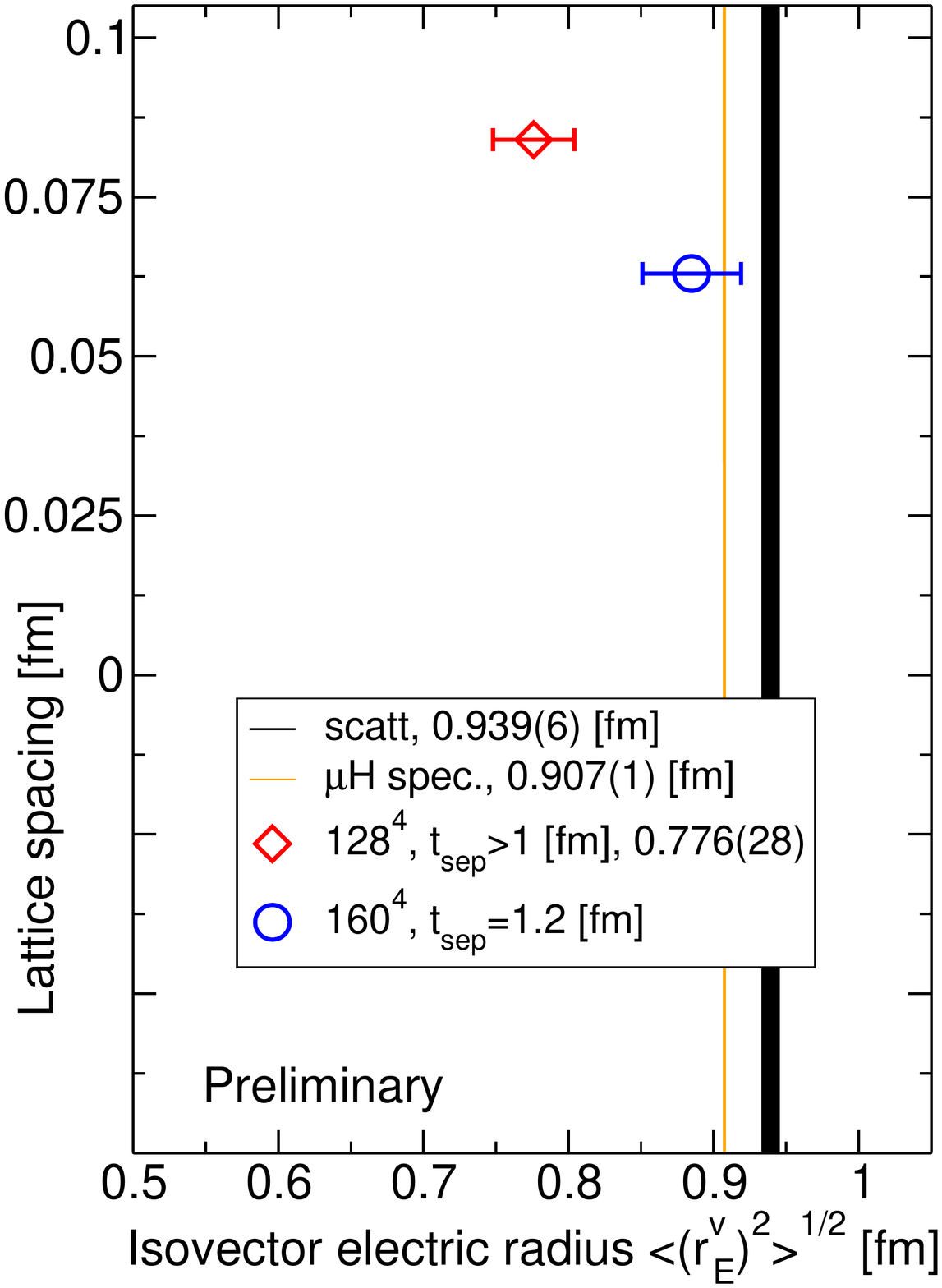}
 \includegraphics[width=0.32\textwidth,bb=10 20 752 725, clip]{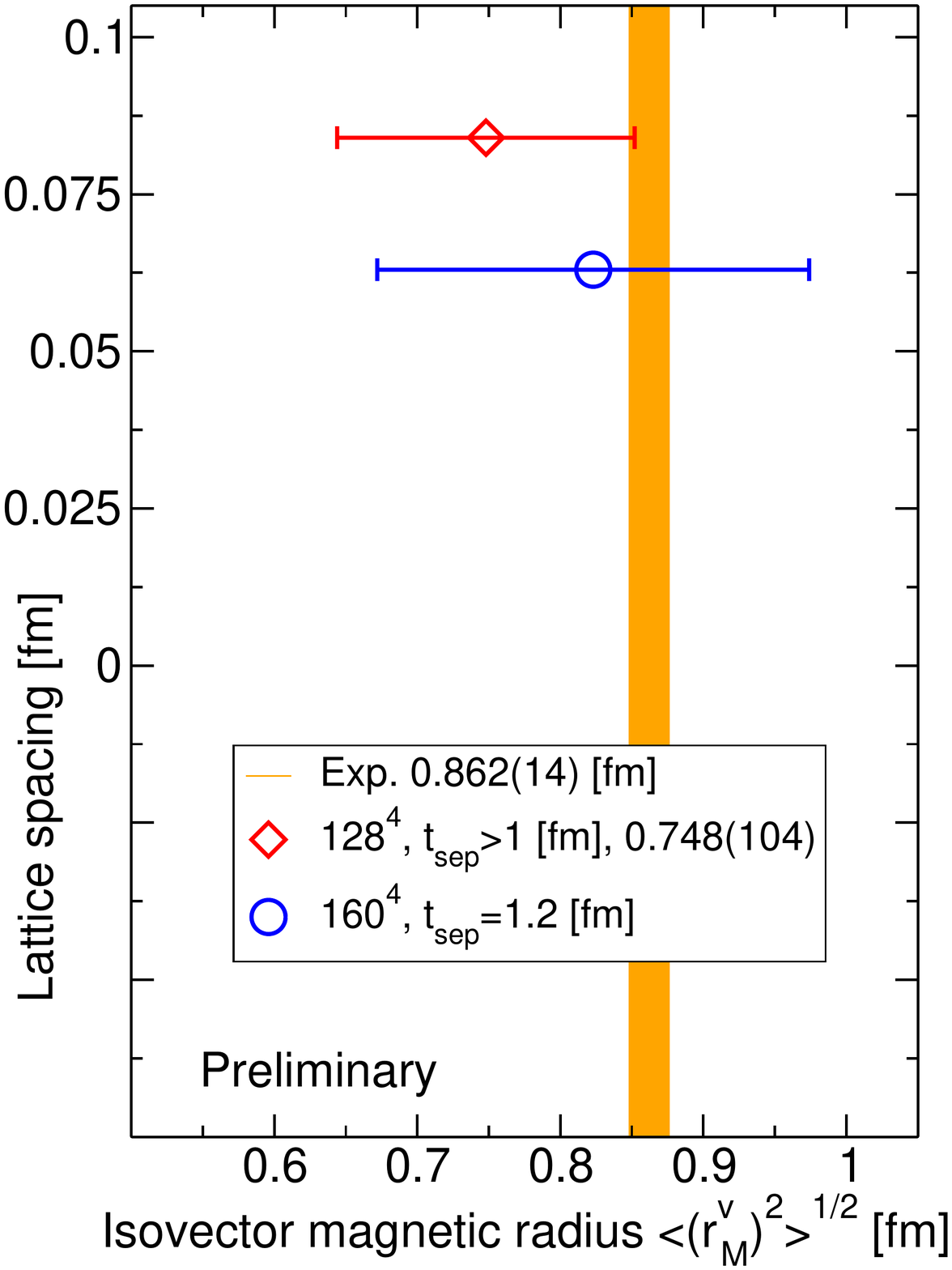}
 \includegraphics[width=0.32\textwidth,bb=10 20 752 725, clip]{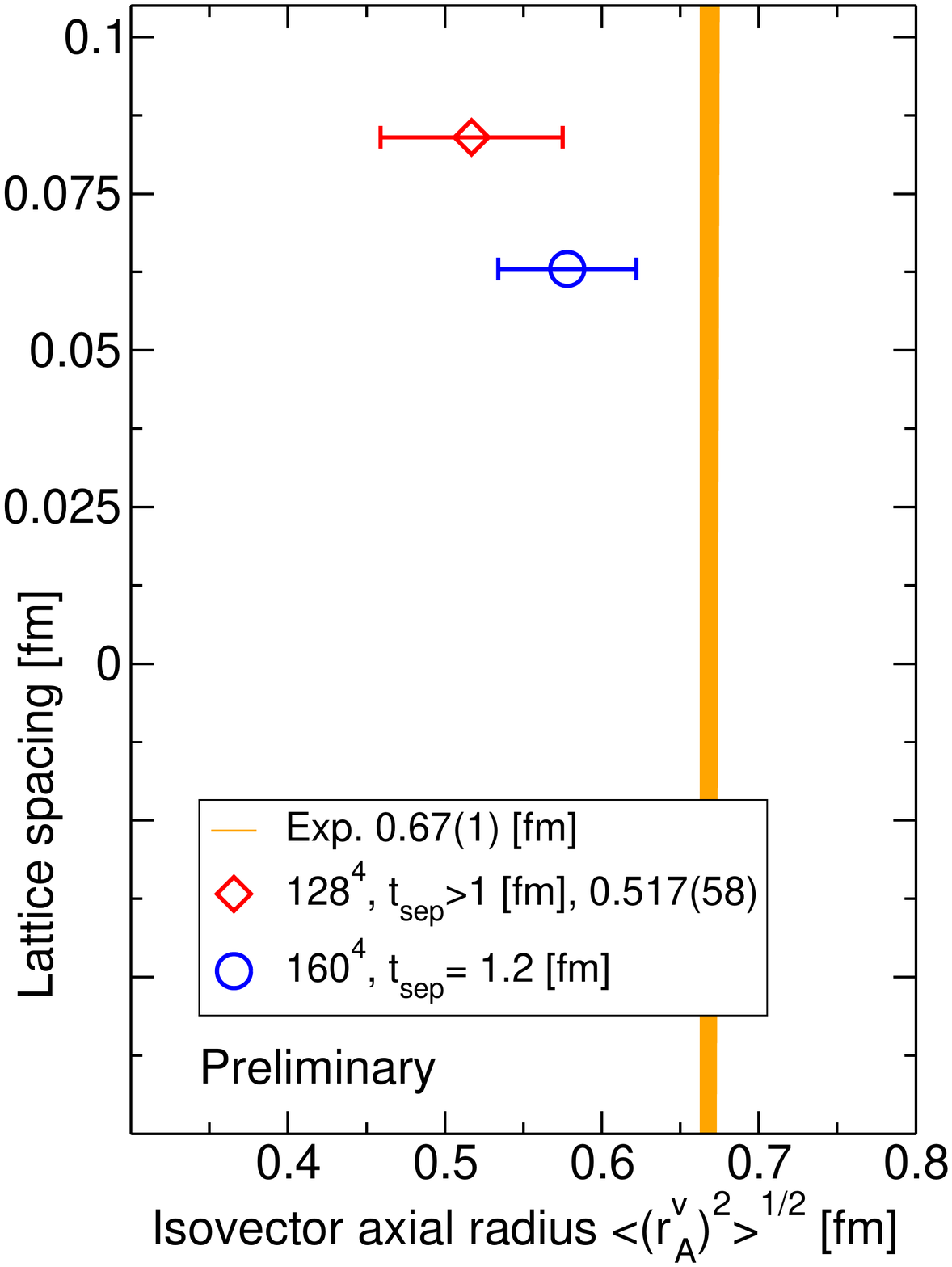}
 \caption{Preliminary results of the isovector electric (left), magnetic (center) and axial (right) RMS radii obtained from the $128^4$ (red diamond) and $160^4$ (blue circle) lattice ensembles. The black and orange bands shown in the isovector electric radius represent the results obtained from lepton-nucleon elastic scattering and muonic hydrogen spectroscopy. An for the other channels, the orange bands represent their experimental or phenomenological values.
}
 \label{fig:continuum_radii}
\end{figure*}

We next present the preliminary results of three kinds of the nucleon isovector RMS radii, such as 
the electric, magnetic and axial RMS radii, obtained from the $160^4$ lattice ensemble, compared with those of the $128^4$ lattice ensemble. 

In three panels of Fig.~\ref{fig:continuum_radii}, we show results of the RMS radii in the three channels. All of these values including the $128^4$ lattice results are given by the z-expansion method~\cite{Boyd:1995cf, Hill:2010yb}, which is the model-independent method in the determination of the nucleon RMS radii from the corresponding form factors. As for the preliminary results of $160^4$ lattice, we only use the data of $t_{\rm sep}/a=19$. This is simply because the larger value of $t_{\rm sep}/a$ yields the smaller systematic uncertainties stemming from the excited-state contaminations. 

As shown in the left panel of Fig.~\ref{fig:continuum_radii}, the discretization uncertainty of the electric RMS radius is evaluated to be more than 10\%, which is much larger than their statistical errors of about 3-4\%. This 
indicates that the $128^4$ lattice result suffers more severely from the discretization uncertainty than the preliminary $160^4$ lattice result. 
Further studies are needed to confirm that the other systematic errors from the excited-state contamination are well controlled to draw a firm conclusion. 
On the other hand, as for the magnetic and axial RMS radii, the sizes of their discretization errors seem to be comparable with their statistical errors. This suggests
that the discretization uncertainties of the magnetic and axial RMS radii, are not large in contrast to the electric one.

\section{Summary}

We have evaluated the appropriate ratio of $g_{A}/g_{V}$ and three types of the isovector RMS radii, such as electric, magnetic and axial ones at two lattice spacings of 0.085 fm 
and 0.063 fm towards the continuum limit, using the PACS10 gauge configurations.
In this work, we examine the discretization uncertainties 
on the nucleon elastic form factors obtained from our lattice QCD calculations in a large spatial extent of about 10 fm at the 
physical point. 
First, we have succeeded in reproducing the experimental value of the renormalized axial charge with both coarse and fine lattice ensembles at the level of the statistical precision of about 2\%. The resultant values are obtained under the condition of $t_{\rm sep}\ge1$ fm, where the systematic uncertainties from excited-state contamination are kept below 2\%.

We have also evaluated three types of the nucleon isovector RMS rasii  such as electric, magnetic and axial ones. 
First, the discretization uncertainty of the electric RMS radius is evaluated by comparing the results
calculated at two lattice spacings, and then we found that there is 
a rather large discretization error on the electric RMS radius
in contrast with the quantity of the renormalized axial charge.
On the other hand, as for the cases of both magnetic and axial radii, the size of the discretization uncertainties  seem to be still comparable with their statistical errors, and thus we do not draw a firm conclusion. In our future projects with the $160^4$ lattice ensemble, we plan to further study the systematic uncertainties stemming from the excited-state contamination 
with other sets of $t_{\mathrm{sep}}$ so as to make sure they are well under control.
Needless to say that additional lattice simulations using {\it the third PACS10 ensemble}~\cite{Kuramashi:2022lat} is required for achieving a comprehensive study of the discretization uncertainties and then taking the continuum limit of our target quantities.

\section*{Acknowledgement}
We would like to thank members of the PACS collaboration for useful discussions. 
R.~T. is supported by the RIKEN Junior Research Associate Program
, and acknowledge the support from Graduate Program on Physics for the Universe (GP-PU) of Tohoku University.
Numerical calculations in this work were performed on Oakforest-PACS in Joint Center for Advanced High Performance Computing (JCAHPC) and Cygnus in Center for Computational Sciences at University of Tsukuba under Multidisciplinary Cooperative Research Program of Center for Computational Sciences, University of Tsukuba, and Wisteria/BDEC-01 in the Information Technology Center, The University of Tokyo. This research also used computational resources of the Supercomputer Fugaku through the HPCI System Research Projects (Project ID: hp170022, hp180051, hp180072, hp180126, hp190025, hp190081, hp200062, hp200188, hp210088, hp220050) provided by Information Technology Center of the University of Tokyo and RIKEN Center for Computational Science (R-CCS). The  calculation employed OpenQCD system (http://luscher.web.cern.ch/luscher/openQCD/). 
This work is supported by the JLDG constructed over the SINET5 of NII.
This work was also supported in part by Grants-in-Aid for Scientific Research from the Ministry of Education, Culture, Sports, Science and Technology (Nos. 18K03605, 19H01892, 22K03612).

\end{document}